\documentstyle[twocolumn,epsfig]{jpsj}

\def\runtitle{Phase Diagram of a 2D Vertex Model}
\def\runauthor{Hiroshi {\sc Takasaki}, Tomotoshi {\sc Nishino}
and Yasuhiro {\sc Hieida}}

\title{Phase Diagram of a 2D Vertex Model}

\author
{ Hiroshi {\sc Takasaki}, Tomotoshi {\sc Nishino} \footnote{E-mail: 
nishino@phys.sci.kobe-u.ac.jp} and Yasuhiro {\sc Hieida} }

\inst
{Department of Physics, Faculty of Science, 
Kobe University, Rokkodai 657-8501\\}

\recdate{ \today }

\kword{Vertex Model, DMRG, CTMRG, Phase Transition}

\begin{document}
\sloppy
\maketitle

The two-dimensional (2D) vertex model is one of the most 
intensively studied systems in statistical physics.~\cite{Bax}
Analytic expressions of thermodynamic and correlation 
functions are known in the parameter area where Yang-Baxter
relation is satisfied.~\cite{Bax,Yang} The phase diagram 
outside the solvable area is, however, not known completely.

As a small step to explore the phase diagram of the 16-vertex 
model, we investigate its subset which allows 7 vertex configurations
shown in Fig.~1, where thick and thin lines, respectively, correspond to 
the spin states $0$ and $1$ on the bond.~\cite{conf}
The Yang-Baxter relation is not satisfied since 
there is no configuration where two $1$s touch such as
$(1100)$ and $(0110)$. 

If we interpret $0$ and $1$, respectively, 
as absence and presence of adsorbed atoms on the solid surface, the 
condition in Fig.~1 is equivalent to setting the minimum distance between 
adsorbed atoms as the lattice constant. In other word, the model is 
equivalent to the hard square model~\cite{Bax} on the diagonal lattice.

\begin{figure}
\epsfxsize=67mm 
\centerline{\epsffile{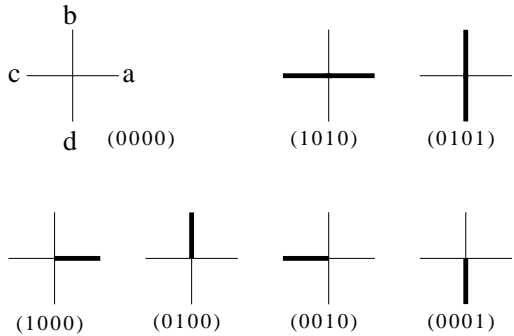}}
\caption{Allowed vertex configurations, that are
$(abcd) =$ $(0000)$, $(1010)$, $(0101)$,
$(1000)$, $(0100)$, $(0010)$,  and 
$(0001)$.}
\label{fig:1}
\end{figure}

An alternative interpretation of the model is, as it has been done for the
6-vertex model, to regard it as 2D lattice polymer on the solid surface,
where $0$ ($1$) represents absence (presence) of a unit monomer on the 
bond. According to the allowed configurations in Fig.~1, only 
straight polymers are allowed to exist. As shown in Fig.~2, it is easily 
imagined that ordered phase appears when the polymer length is long 
or its density is high, and  that disordered phase appears under the 
opposite conditions. The aim of this short note is to draw the phase 
boundary of this order-disorder transition. 

\begin{figure}
\epsfxsize=72mm 
\centerline{\epsffile{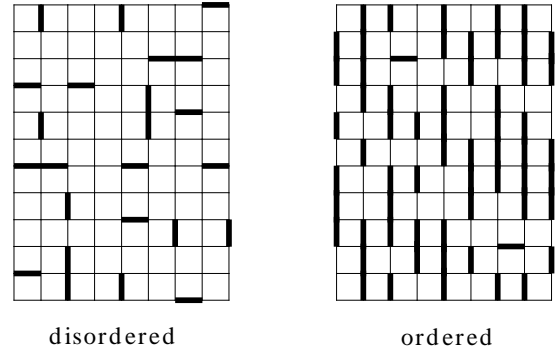}}
\caption{Disordered state and the ordered state.}
\label{fig:2}
\end{figure}

Let us parameterize the vertex model from the view point of the lattice
polymer. For simplicity,  we assume that the local Boltzmann weights are 
symmetric under the right angle rotations. Each vertex weight is then
expressed as
\begin{eqnarray}
W(0000) &=& \exp( 0 ) = 1 \nonumber\\
W(1010) = W(0101) &=& \exp( K ) \nonumber\\
W(1000) = W(0100) = W(0010) && \\
= W(0001) &=& \exp( K/2 + B ) \nonumber
\end{eqnarray}
using two parameters $K \equiv -\beta U$ and $B \equiv -\beta V$, 
where $U$ is the energy per length of the line polymer and $V$ is 
its boundary energy; the Boltzmann weight for a straight line of 
length $N$ is $\exp( NK + 2B )$. 

An appropriate order parameter for the order-disorder transition in 
Fig.~2 is
\begin{eqnarray}
M &=& P(1010) + \frac{P(1000) + P(0010)}{2} \nonumber\\
    &-& P(0101) - \frac{P(0100) + P(0001)}{2} \, ,
\end{eqnarray}
where $P(abcd)$ is the absolute probability to observe the local 
configuration $(abcd)$. We calculate $M$ at the center of 
square systems with linear dimension $L$ using the CTMRG
method,~\cite{CTMRG} which is a variant of the density matrix
renormlaization group (DMRG)~\cite{Wh1,Dresden} applied
to 2D classical systems,~\cite{Ni} where $L$ 
is chosen to be sufficiently larger than the correlation length. 
To weakly stabilize the order in the horizontal direction, we impose 
fixed boundary conditions where all the spins on the vertical sides 
of the square are $1$ and those on the horizontal sides are $0$.
This boundary condition makes the corner transfer matrix 
asymmetric, therefore we use an asymmetric extension of 
the CTMRG method.

\begin{figure}
\epsfxsize=67mm 
\centerline{\epsffile{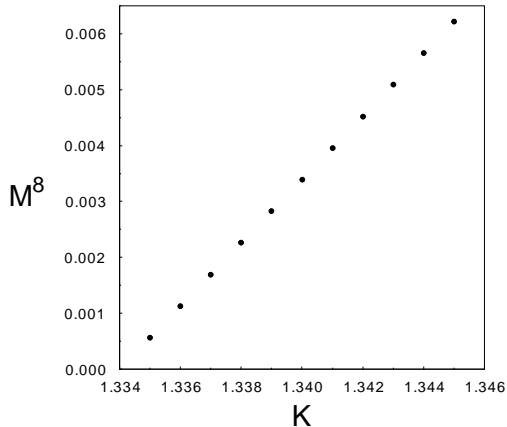}}
\caption{The eighth power of the order parameter ($= M^8_{~}$) 
with respect to $K$ when $B = 0$.}
\label{fig:3}
\end{figure}

The phase boundary is determined from the observation of the 
order parameter $M$. Figure 3 shows the $K$-dependence of 
$M^8_{~}$ when $B = 0$. It is clear that $M^8_{~}$ is a linear
function of $K$, and we obtain the critical point 
$K_{\rm c}^{~} = 1.334$. Also for the cases where $B \neq 0$
we have checked that $M^8_{~}$ is linear in $K$; 
the critical index $\beta$ of this system is $1/8$. 

\begin{figure}
\epsfxsize=67mm 
\centerline{\epsffile{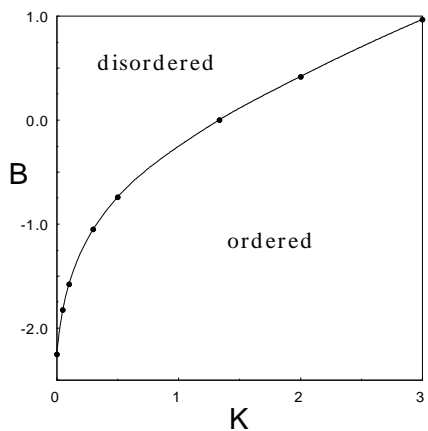}}
\caption{Phase diagram in the $K$-$B$ plane.}
\label{fig:4}
\end{figure}

We thus obtained the phase boundary in the $K$-$B$ plane shown in 
Fig.~4. The system is always disordered in the negative $K$ region, therefore 
we only show the positive $K$ region. The boundary starts from 
$( K, B ) = ( 0_{+}^{~}, -2.25 )$, it passes $( K, B ) = ( 1.334, 0 )$, 
and then approaches the line $B = ( K - 1.0 ) / 2$; 
we have not identified the reason why the line has the offset of amount
$- 1.0/2 = -0.5$. 

It is worth obtaining another critical index of this model, which is related
to the density defined by
\begin{eqnarray}
n &=& P(1010) + \frac{P(1000) + P(0010)}{2} \nonumber\\
    &+& P(0101) + \frac{P(0100) + P(0001)}{2} \, .
\end{eqnarray}
Figure 5 shows the $K$ dependence of $n - n_{\rm c}^{~}$ when 
$B = 0$, where $n_{\rm c}^{~} = 0.736$ is the density at the
phase boundary $K_{\rm c}^{~} = 1.334$. It is apparent that
$n - n_{\rm c}^{~}$ is proportional to $K - K_{\rm c}^{~}$. Since
$n$ is related to the energy parameter $U$, the exponent $\alpha$ 
is equal to $0$. 

\begin{figure}
\epsfxsize=67mm 
\centerline{\epsffile{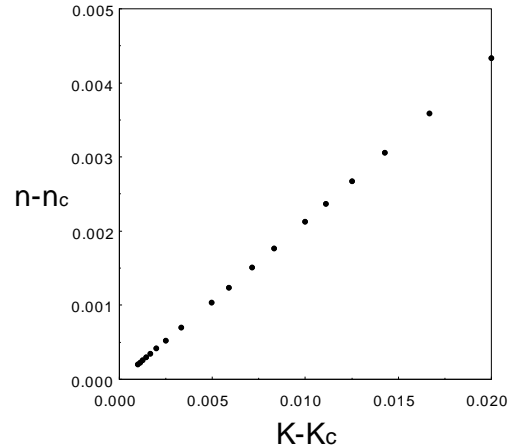}}
\caption{Density in Eq.~3 near the transition point 
$K_{\rm c}^{~} = 1.334$ when $B = 0$, where $n_{\rm c}^{~} = 0.736$.}
\label{fig:5}
\end{figure}

In conclusion we have studied a symmetric vertex model which 
allows 7 vertex configurations shown in Fig.~1, and obtained its 
phase diagram. The critical indices of this model, that are 
$\beta = 1/8$ and $\alpha = 0$, are the same as those of the 
Ising model. To include additional configurations, such as $(1111)$ and
$(1110)$, that may cause the percolation transition, is our
next subject toward the complete understanding of the 16-vertex
model.

We thank to T.~Hikihara and Y.~Akutsu for discussions about DMRG.
T.~N. thank to  G.~Sierra and M.A.~Mart\'{\i}n-Delgado for valuable
discussions at C.S.I.C. He also thank to A.~Gendiar.
The present work is partially supported by a Grant-in-Aid 
from Ministry of Education, Science and Culture of Japan
(No. 11640376) and also JSPS Research Fellowship for
young scientists.


\begin{thebibliography}{99}
\bibitem{conf} It should be noted that the choice of 7 vertices is different from 
that of the so called the 7-vertex model.
\end{thebibliography}
\end{document}